\documentclass[12pt]{article}
\usepackage[utf8]{inputenc}
\usepackage[T1]{fontenc}
\usepackage{lmodern, dsfont, pifont}
\usepackage[round, sort]{natbib}
\usepackage[french, english]{babel}
\usepackage{xcolor,graphicx,psfrag}
\definecolor{Blue}{rgb}{0.1,0.1,0.5}
\usepackage[hypertexnames=false]{hyperref}
\hypersetup{bookmarksnumbered,
            colorlinks=true,
            citecolor=Blue,
            linkcolor=black,
            breaklinks,
            pdfborder=0 0 0,
            pdftitle=,
            pdfauthor=,
            pdfkeywords=}

\usepackage{amsmath,amsthm,amsfonts,amssymb,bm}
\theoremstyle{remark}
\newtheorem{remark}{Remark}

\usepackage[linesnumbered,ruled,vlined]{algorithm2e}
\SetKwInput{KwInput}{Input}
\SetKwInput{KwOutput}{Output}

\newcommand   \du      {\mathrm{d}u}
\newcommand   \Esp     {\mathbb{E}}
\newcommand   \one     {\mathds{1}}
\newcommand   \Rset    {\mathbb{R}}

\newcommand   \Bcal    {\mathcal{B}}
\newcommand   \Lcal    {\mathcal{L}}
\newcommand   \Ncal    {\mathcal{N}}
\newcommand   \Ocal    {\mathcal{O}}
\newcommand   \Pcal    {\mathcal{P}}

\newcommand   \Jphi    {J_{\varphi, w}}
\newcommand   \tra     {{\text{\raisebox{1pt}{\tiny T}}}}

\begin{document}

\begin{center}
  \Large\linespread{1.0}\textsc{%
    Towards new cross-validation-based estimators %
    for Gaussian process regression:\\
    efficient adjoint computation of gradients%
  }

\normalsize

\bigskip

S\'ebastien J. Petit$^{1, 2, \star}$ \& Julien Bect$^{1}$ \& S\'ebastien Da
Veiga$^{3}$ \\
\& Paul Feliot$^{2}$ \& Emmanuel Vazquez$^{1}$ \bigskip

{\it \selectlanguage{french} %
  $^{1}$ Universit\'e Paris-Saclay, CNRS, CentraleSupélec,
  Laboratoire des Signaux et Systèmes, Gif-sur-Yvette,
  France. $^\star$E-mail: \texttt{sebastien.petit@centralesupelec.fr}.

  $^{2}$ Safran Aircraft Engines, Moissy-Cramayel, France \\

  $^{3}$ Safran Tech, Ch\^ateaufort, France}
\end{center}
\bigskip

\selectlanguage{french} {\bf R\'esum\'e.} Nous nous intéressons à
l'estimation par validation croisée des paramètres d'une fonction de
covariance d'un processus gaussien. Nous suggérons l'utilisation de
nouveaux critères de validation croisée dérivés de la littérature des
\textit{scoring rules}.  Nous proposons de plus une méthode efficace
pour le calcul du gradient d'un critère de validation.  Cette méthode est plus
efficace que ce qui est présenté dans la littérature à notre
connaissance, et permet en particulier de réduire la complexité de
l'évaluation jointe des critères de validation croisée et des gradients
associés.

{\bf Mots-cl\'es.} Processus gaussien, validation croisée,
score de prédiction probabiliste, Mode adjoint

\medskip \selectlanguage{english} {\bf Abstract.}  We consider the
problem of estimating the parameters of the covariance function of a 
Gaussian process by cross-validation. We suggest using new
cross-validation criteria derived from the literature of \textit{scoring
  rules}.  We also provide an efficient method for computing the gradient
of a cross-validation criterion.
To the best of our knowledge, our method is more efficient than what has
been proposed in the literature so far. It makes it possible to lower the
complexity of jointly evaluating leave-one-out criteria and their
gradients.


\bigskip\bigskip


\section{Introduction}
\label{sec:intro}

Let $\xi$ be a zero-mean Gaussian process indexed by $\Rset^d$ and denote
by $k$ the covariance function of $\xi$, which is assumed to belong to a
parametrized family $\{k_{\theta} ;{\theta \in \Theta} \}$, where
$\Theta \subseteq \Rset^q$ denotes a $q$-dimensional space of parameters.
We can safely say that the most popular methods for estimating $k$ from data are maximum
likelihood and related techniques.

In this article we focus instead on cross-validation methods.
Classical cross-validation methods for estimating $k$ are based on the
leave-one-out mean squared prediction error or PRESS
\citep{allen1974relationship, bachoc-gcv-mle}, and leave-one-out
log predictive density \citep[see, e.g.,][]{Rasmussen}.
These leave-one-out goodness-of-fit criteria can be computed using
closed-form formulas \citep{Dubrule1983CrossVO}.

The contribution of this work is twofold.
First, we suggest extending the range of classical cross-validation
criteria available in the literature of Gaussian processes by using
the broad variety of \textit{scoring rules} \citep[see][for a
survey]{Gneiting}, such as the continuous ranked probability score
(CRPS).
Second, we provide an efficient way for computing the gradient of any
cross-validation criterion, which can then be used in gradient-based
optimization algorithms.
The only requirement is for the criterion to be differentiable in
closed form with respect to leave-one-out posterior predictive means
and variances.
The new procedure has a $\mathcal{O}(n^3 + qn^2)$ complexity, against the
$\mathcal{O}(qn^3)$ that was deemed ``unavoidable'' by
\cite{Rasmussen}.

The article is organized as follows.
Section~\ref{sec:scoring} introduces scoring rules and how they can be
used for estimating $k$.
Section~\ref{sec:reverse} presents the details of our contribution to
the computation of gradients of a cross-validation criterion
and Section~\ref{sec:concl} presents our conclusions and perspectives.

\section{Scoring rules and cross-validation criteria}
\label{sec:scoring}

Let $Z_i = \xi(x_i) + \varepsilon_i$,
$1 \leq i \leq n$, be some observations of $\xi$, at
points $x_i\in\Rset^d$, where the $\varepsilon_i$s are assumed independent and identically
$\Ncal(0, \sigma_\varepsilon^2)$-distributed, with $\sigma_{\varepsilon}^2 \geq 0$. 

The classical framework of Gaussian process regression allows one
to build a predictive distribution for an unobserved
$\xi(x)$ at $x \in \Rset^{d}$ from the $Z_i$s.
Criteria for assessing the quality of probabilistic predictions have been studied in depth
under the name of \textit{scoring rules} in the seminal article of
\cite{Gneiting}.
A scoring rule for real variable prediction is a function
$S : \Pcal\, \times\, \Rset \to \left[-\infty, +\infty\right]$, where $\Pcal$
is a class of probability measures on $(\Rset, \Bcal(\Rset))$.
For $P \in \Pcal$ and $z \in \Rset$, $S(P, z)$ measures the goodness of
prediction $P$ for $z$.

Assume that we want to use a scoring rule $S$ for estimating the
parameters of a covariance function.
\cite{Gneiting} suggest building a leave-one-out
cross-validation criterion $L$ defined as
\begin{equation}\label{LOO_cross_validated_criterion}
  L( \theta) = \frac{1}{n} \sum_{i = 1}^n S(P_{-i}^{\theta}, Z_i),
\end{equation}
where $P_{-i}^{\theta}$ is the conditional distribution of $Z_{i}$ given
the $Z_j$s, for $\ 1 \leq j \leq n, \ j \neq i$.

In our Gaussian process regression framework, it is well known that
$P_{-i}^{\theta}$ is a Gaussian distribution $\Ncal(\mu_i, \sigma_i^2)$.  Let
$K = (k_\theta(x_i, x_j))_{1 \leq i,j \leq n}$ be the covariance matrix
of $(\xi(x_1),\ldots,\xi(x_n))^\tra$, then \citep{Dubrule1983CrossVO, sunda:2001,
  Craven1978SMOOTHINGND} show that 
\begin{equation}\label{LOO_formulas}
  \mu_i = Z_i - \frac{(B Z)_i}{B_{i, i}} \quad \mathrm{and}
  \quad \sigma_i^2 = \frac{1}{B_{i, i}},
\end{equation}
where $B = (K + \sigma_\varepsilon^2 I)^{-1}$ and
$Z=(Z_1,\ldots,Z_n)^\tra$. Note that \eqref{LOO_formulas} still stands
true if $\sigma_\varepsilon^2 = 0$.

\begin{remark}\label{generalization}%
  \citet[Lemma 3.1 and 3.2]{Craven1978SMOOTHINGND} show
  that~\eqref{LOO_formulas} could be generalized to other types of linear
  predictors, beyond the particular Gaussian process regression framework
  considered in this article.
\end{remark}

The mean squared prediction error and log-predictive density
criteria mentioned in Section~\ref{sec:intro} correspond
respectively to the scoring rules
$S_1(P, z) = -(\Esp_{Z \sim P}(Z) - z)^2$ and $S_2(P, z) = \log(f(z))$,
where $f$ denotes the density of $P$ with respect to some reference measure.
A scoring rule is said \textit{strictly proper} if
$\Esp_{Z \sim P}(S(P, Z)) > \Esp_{Z \sim P}(S(Q, Z))$ for all
$P,~Q \in \Pcal$ with $P \neq Q$.
Strict propriety can be viewed as a sanity condition for performing
estimation by maximizing \eqref{LOO_cross_validated_criterion}.
Note that $S_1$ is not strictly proper relative to the class of
Gaussian measures whereas $S_2$ is.
A large variety of scoring rules is surveyed by \cite{Gneiting}.
We shall use the CRPS in Section~\ref{sec:concl} for illustration.

\section{Efficient computation of the gradient of a leave-one-out criterion}
\label{sec:reverse}

In this section we present our contribution for computing the gradient
$\nabla_{\theta} L$ of~(\ref{LOO_cross_validated_criterion}). Let\footnote{%
  We identify the space of $n \times n$ matrices with $\Rset^{n^2}$ and
  ${(\Rset^n)}^2$ with $\Rset^{2n}$ with a slight abuse of notation.}

\begin{equation}
  \left\{
    \begin{array}{l}
      \Gamma : \theta \in \Rset^{q} \mapsto K \in \Rset^{n^2},\\
      \varrho : K \in \Rset^{n^2} \mapsto (\mu, \sigma^2) \in \Rset^{2n} \ \mathrm{according \ to} \  \eqref{LOO_formulas},\\
      \varphi: (\mu, \sigma^2) \in \Rset^{2n} \mapsto L \in \Rset \ \mathrm{according \ to} \
      (\ref{LOO_cross_validated_criterion}),\\
    \end{array}
  \right.
\end{equation}
where $\mu = (\mu_1, ..., \mu_n)^\tra$ and
$\sigma^2 = (\sigma_1^2, ..., \sigma_n^2)^\tra$,
in such a way that
$L(\theta) = \left( \varphi \circ \varrho \circ
  \Gamma\right)(\theta)$.
Write $w = (\mu, \sigma^2)$ for simplicity.
Let $\Jphi$, $J_{\varrho, K}$ and $J_{\Gamma, \theta}$ be the
$1 \times 2n$, $2n \times n^2$, $n^2 \times q$ Jacobian matrices of
$\varphi$, $\varrho$ and $\Gamma$ at  $w$, $K$
and $\theta$ respectively.
Using the chain rule for derivation we have
\begin{equation}\label{chain_rule}
  \nabla_{\theta} L^\tra = \Jphi\, J_{\varrho, K}\, J_{\Gamma, \theta}.
\end{equation}
\cite{Rasmussen} propose an algorithm in $\Ocal(qn^3)$ time for
computing $\nabla_{\theta} L$ from $J_{\Gamma, \theta}$.

Suppose that these Jacobian matrices are already built and stored.
Then, computing~\eqref{chain_rule} by multiplying those matrices from
the right to the left costs about
$2n \cdot n^2 \cdot q + 1 \cdot 2n \cdot q = \Ocal(qn^3)$ additions
and multiplications, corresponding to the complexity announced by
\cite{Rasmussen}.
On the other hand, proceeding from the left to the right costs about
$1 \cdot 2n \cdot n^2 + 1 \cdot n^2 \cdot q = 2n^3 + qn^2$ additions
and multiplications.
\citep[This kind of consideration is a basic illustration of what has
been studied in depth in the literature as the \textit{matrix chain
  multiplication} problem for variable length products of matrices;
see, e.g.,][and references therein.]{hu-shing:1982}

Let us now investigate the price paid for building $\Jphi$ and
$J_{\varrho, K}$.
First of all, the computation of $B = (K + \sigma_\varepsilon^2 I)^{-1}$ and then
$w = (\mu, \sigma^2)$ from $K$ can be performed in $\Ocal(n^3)$
operations using~\eqref{LOO_formulas}.
Moreover, knowing $w$, $L$ and $\Jphi$ can be computed in $\Ocal(n)$ time.
In addition, equations used by \cite{sunda:2001} show that
$J_{\varrho, K}$ can be build from $B$ in $\Ocal(n^3)$ elementary
operations.
Thus, previous arguments show that it is indeed possible to compute
$L$ and $\nabla_{\theta} L$ from $J_{\Gamma, \theta}$ and $K$ for
$\Ocal(n^3 + qn^2)$ elementary operations, thereby avoiding the
$\Ocal(qn^3)$ complexity mentioned by \cite{Rasmussen}.

Furthermore, available implementations \citep[see, e.g.,][]{STK} show
that it is possible build $K$ and  $J_{\Gamma, \theta}$ from
$\theta$ in a $\Ocal(qn^2)$ complexity for the case of an anisotropic
stationary covariance with $q = d +1$ parameters (one variance
parameter and $q$~length scales).
We see then that our contribution allows us in this case to keep the
evaluation of $L$ and $\nabla_{\theta} L$ from $\theta$ in
$\Ocal(n^3 + qn^2)$, rather than $\Ocal(qn^3)$.

The main drawback of this scheme is the $2n \times n^2$ storage of
$J_{\varrho, K}$. We propose to circumvent this cost by directly
implementing the adjoint operators of the differentials of~$\varrho$:
\begin{equation}
  \Lcal_\varrho^*: \left( K, \delta_w \right)
  \mapsto J_{\varrho, K}^\tra\, \delta_w.
\end{equation}
This can be used to compute
$\Lcal_\varrho^*(K, \Jphi^\tra) = J_{\varrho, K}^\tra \Jphi^\tra$ and
then $\nabla_{\theta} L$ from \eqref{chain_rule}.  This way of
implementing chain rule derivatives is well known and has been studied
under the name of \textit{reverse-mode differentiation}\footnote{%
  In the context of Gaussian process regression, a reverse-mode
  differentiation approach has been proposed by \cite{Toal} for the
  computation of the likelihood function and its gradient.} %
or \textit{backpropagation}, and its paternity can be traced back at
least to \cite{Linnainmaa:1970}.

We propose Algorithm~\ref{dK_from_dmu_loo_dvar_loo} to implement this
operator.
This algorithm only requires $\Ocal(n^2)$ storage capacity and about
$2n^3$ additions and multiplications,
thus reducing the burden of storage, while maintaining the global
$\Ocal(n^3 + qn^2)$ complexity.
(Note that $2n^3$ already corresponds to the cost of matrix
multiplication in a ``direct'' approach that would first
build~$J_{\varrho, K}$ and then compute
$\Lcal_\varrho^*\left( K, \delta_w \right)$ by matrix multiplication.)

\begin{algorithm}[t]\label{dK_from_dmu_loo_dvar_loo} \small
  \KwInput{\\
    $K$, $Z$, $B = (K + \sigma_\varepsilon^2 I)^{-1}$,
    $ \alpha = By$,
    $ \kappa = (B_{i,i})_{1 \leq i \leq n}$,
    $ \kappa^{-1} = \one \oslash \kappa$,
    $ \chi =  \alpha \circ  \kappa^{-1}$,
    $ \delta_w = (\delta_\mu, \delta_{\sigma^2})$}
  \KwOutput{$\delta_{K} = J_{\varrho, K}^\tra \delta_w$}
  $ \kappa^{-2} =  \kappa^{-1} \circ \kappa^{-1}$\\
  $ \delta_\chi = - \delta_\mu$\\
  $ \delta_\alpha = \delta_\chi \circ \kappa^{-1}$\\
  $ \delta_\kappa = - \delta_\chi \circ {\alpha} \circ \kappa^{-2} - \delta_{\sigma^2} \circ \kappa^{-2}$\\
  $ \delta_B = \delta_\alpha Z^\tra + \mathrm{diag}(\delta_\kappa)$\\
  $ \delta_{K} = - B^\tra\, \delta_B\, B^\tra$\\
  \caption{\small %
    Implementation of $\Lcal_\varrho^*$ for computing
    $\delta_{K} = J_{\varrho, K}^\tra \delta_w$,
    from $K$ and $\delta_w$.
    Inputs at first
    step refer to what
    has already been
    computed for evaluating
    $\mu$ and $\sigma^2$. For
    vectors $a$ and $b$,
    $a \oslash b$ and $a  \circ b$ denote the Hadamard
    element-wise division and multiplication
    respectively.}
\end{algorithm}

\medbreak

\begin{remark}
  The algorithm can easily be adapted, through a suitable modification
  of the matrix~$B$ used in Step~6, to any type of linear model for
  which~\eqref{LOO_formulas} holds (see Remark~\ref{generalization}).
\end{remark}

\vspace{-5mm}

\section{Conclusion and perspectives}
\label{sec:concl}

We suggested using the scoring rules referenced by \cite{Gneiting} for
the estimation of the parameters of a 
Gaussian process by leave-one-out cross-validation.
We also proposed an efficient procedure for computing gradients of
cross-validation criteria that is more efficient than what was
available in the literature to our knowledge.

\begin{figure}
  \begin{center}
    \psfrag{rho1}{$\rho_1$}
    \psfrag{rho2}[Bc][Bc][1][180]{$\rho_2$}
    \includegraphics[width=9.5cm]{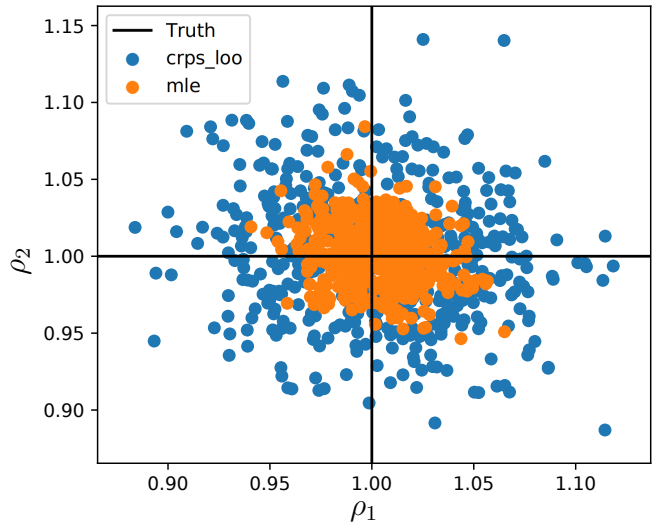}
  \end{center}
  \vspace{-9mm}
  \caption{%
    Scatterplots of the  estimates of the two length scales of a
    Gaussian process on~$\Rset^{2}$.
    Blue points correspond to CRPS-based cross-validation estimates;
    orange points correspond to maximum likelihood estimates.
    True length scales are represented by black lines. Each scatterplot
    consists of $500$ estimations obtained from a space filing design of
    size $n=500$.
    The criteria were optimized using a quasi-Newton type algorithm.
  }
\label{clouds}
\end{figure}

Further work will consist in investigating the properties of these
estimators for several scoring rules.
For instance, one can choose to use the continous rank probability
score (CRPS) defined as
$\mathrm{CRPS}(F, z) = -\int_{-\infty}^{+\infty} \left(F(u)- \one_{z
    \leq u} \right)^2 \du$, where $F$ is a cumulative distribution
function.
The CRPS is strictly proper relative to the class of Gaussian
measures\footnote{and more generally with respect to the class of all
  probability measures with finite first order moment \citep[see,
  e.g.][Section~4.2]{Gneiting}}.
An empirical comparison with maximum likelihood for estimating the length scales
is presented in Figure~\ref{clouds}.
Our contribution for computing gradients makes it possible to maintain
the same complexity, both in terms of storage and calculation, for the
two methods.

\bibliographystyle{unsrtnat}
\bibliography{jds2020-spetit.bib}

\end{document}